\documentclass[useAMS,usenatbib,usegraphicx]{mn2e}

\newcommand{\etal}{et al.\ }

\newcommand{\beq}{\begin{equation}}
\newcommand{\beqa}{\begin{eqnarray}}
\newcommand{\eeq}{\end{equation}}
\newcommand{\eeqa}{\end{eqnarray}}

\title[Detecting Reionization in the Star Formation Histories of
High-Redshift Galaxies] {Detecting Reionization in the Star Formation
Histories of High-Redshift Galaxies}

\author[R. Barkana and A. Loeb]{Rennan Barkana$^{1}$ and Abraham Loeb$^{2}$
\thanks{E-mail: barkana@wise.tau.ac.il (RB); aloeb@cfa.harvard.edu (AL)}\\
$^{1}$School of Physics and Astronomy, The Raymond and Beverly Sackler
Faculty of Exact Sciences,\\ Tel Aviv University, Tel Aviv 69978,
ISRAEL\\ $^{2}$Astronomy Department, Harvard University, 60 Garden
Street, Cambridge, MA 02138, USA}

\begin{document}

\pagerange{\pageref{firstpage}--\pageref{lastpage}} \pubyear{2005}

\maketitle

\label{firstpage}

\begin{abstract}

The reionization of cosmic hydrogen, left over from the big bang,
increased its temperature to $\ga 10^4$K. This photo-heating resulted
in an increase of the minimum mass of galaxies and hence a suppression
of the cosmic star formation rate. The affected population of dwarf
galaxies included the progenitors of massive galaxies that formed
later. We show that a massive galaxy at a redshift $z\ga 6$ should
show a double-peaked star formation history marked by a clear
break. This break reflects the suppression signature from reionization
of the region in which the galaxy was assembled. Since massive
galaxies originate in overdense regions where cosmic evolution is
accelerated, their environment reionizes earlier than the rest of the
universe. For a galaxy of $\sim 10^{12}M_\odot$ in stars at a redshift
of $z\sim 6.5$, the star formation rate should typically be suppressed
at a redshift $z\ga 10$ since the rest of the universe is known to
have reionized by $z\ga 6.5$. Indeed, this is inferred to be the case
for HUDF-JD2, a massive galaxy which is potentially at $z\sim 6.5$ but
is inferred to have formed the bulk of its $3\times 10^{11}M_\odot$ in
stars at $z\ga 9$.

\end{abstract}


\section{Introduction}

Following the epoch of cosmic recombination at $z\sim 10^3$, neutral
hydrogen pervaded the universe until it was reionized by the first
galaxies \citep{Review}. There has been much interest recently in
developing observational techniques that will reveal the redshift of
reionization, $z_{\rm rei}$. Existing methods provide conflicting
constraints. The large-scale polarization anisotropies of the cosmic
microwave background (CMB) imply $11\la z_r\la 30$ at 95\% confidence
\citep{Ko}, while the size of the ionized regions around quasars at
$z \sim 6$ \citep{WL04,MH,WLC} imply a substantial neutral hydrogen
fraction at that late time. Here we propose a new probe of $z_r$,
namely the imprint of reionization on the star formation histories
(SFHs) of individual massive galaxies at $z\ga 6$.

The photo-ionization of hydrogen raised its temperature to $\ga
10^4$K. This heating process resulted in an increase of the minimum
mass of galaxies and hence a suppression of the cosmic star formation
rate (SFR). We previously suggested \citep{SFRcut} that this global
signature could be observed by detecting the corresponding fall in the
number counts of faint galaxies. However, we predicted that even the
future James Webb Space Telescope could individually detect only a
fraction of the affected low-mass galaxies. While the same signature
may be more easily detectable through the evolution of the supernova
rate \citep{mjh05}, there is a generic reason that makes it
advantageous to search for localized rather than global signatures of
reionization. At high redshift, large fluctuations in the number
density of galaxies in various regions causes reionization to occur at
different times in different regions \citep{BLflucts}. Thus, a region
of size $\la 10$ comoving Mpc should show evidence for a relatively
rapid reionization, while any global signature will be further
smoothed out over a redshift interval of $\sim 2$--4.

Since a $10^{12} M_{\odot}$ halo forms out of matter in a region of a
few Mpc in size, the SFH of an individual massive galaxy at high
redshift should show a break in its SFR. This break reflects the
suppression signature from reionization of the region from where the
galaxy was assembled. We show that even a rather massive galaxy at $z
\sim 6$ should show this break, since most of its mass is expected to
have been in low-mass progenitors when the surrounding region was
reionized. Moreover, if this signature can be detected in multiple
galaxies which lie in regions that reionized at different times, then
the spatial variation of reionization can be probed with this method.

Massive galaxies are routinely being detected at redshifts $z\ga 6$
\citep{b05,stark}, in some cases with sufficient data for a detailed
analysis of the stellar content. For example, \citet{z6p5a} have
recently reported the discovery of HUDF-JD2, a galaxy which is
potentially at $z\sim 6.5$, although spectral lines were not detected
and thus the redshift is based on the shape of the spectrum. If the
high redshift is a correct inference, then the galaxy possesses almost
$10^{12}M_\odot$ in stars. Further analysis of the spectral shape, in
particular the Balmer break signature of an established stellar
population, indicates that the galaxy formed the bulk of its stars at
$z\ga9$ and could have reionized its surrounding region at $z\ga10$
\citep{z6p5b}. Also, \citet{eyles05} analyzed two galaxies at $z \sim
5.8$ that had been found by \citet{sbm03}. They showed that these
galaxies had stellar masses of $\ga 10^{10} M_{\odot}$ and had formed
most of their stars at $z \ga 7.5$ while a minority formed closer to
the detected redshift [see also \citet{dickinson}].

It is therefore particularly timely to examine the question of whether
the above-mentioned data imply that reionization occurred early in the
universe. In this paper we consider the imprint of reionization on the
SFHs of individual massive galaxies like HUDF-JD2. Since massive
galaxies originate in overdense regions where cosmic evolution is
accelerated, we show that their environment reionizes earlier than the
rest of the universe. In \S~2 we present our formalism for calculating
the SFH within galaxies and show general results in
\S~3. We apply these results to HUDF-JD2 and the other galaxies in
\S~4 and give our main conclusions.  Throughout the paper, we assume a
flat universe with cosmological parameters $h=0.72$ and
$\Omega_m=0.27$ (which includes an $\Omega_b=0.046$), with the
remaining energy density in a cosmological constant \citep{CMB}.

\section{Modeling Star Formation Histories}

In order to model the SFH of a typical galaxy, we must account for a number
of effects that determine the availability of gas for star formation. We
first consider the ability of gas in general to cool, and the ability of
pressure to resist gravity -- particularly after the gas is heated during
reionization. Having analyzed these effects on the gas, we begin for each
galaxy with the average accretion history of the dark matter halo that will
eventually host it, and then calculate the additional effects that modify
the accretion history of the gas relative to the dark matter. We then
consider how the reionization history differs from region to region, and
how this is correlated with the SFHs of galaxies in different regions.

We begin by considering the gas in galactic halos. Star formation
requires the gas to cool and collapse to high density within a
halo. We assume that galaxies form in halos above some minimum mass
$M_{\rm cool}$ that depends on the efficiency of atomic and molecular
cooling. A minimum virial temperature of $\sim 10^4$~K is required for
significant ionization and thus efficient atomic cooling, and this
imposes a minimum halo circular velocity of 16.5 km/s. On the other
hand, if molecular hydrogen is present, then $T_{\rm vir} \sim 10^3$~K
is needed to excite its rotational and vibrational transitions, which
yields $V_c = 3.7$ km/s for molecular hydrogen cooling. Since feedback
from stellar radiation may efficiently dissociate the $H_2$ in the
intergalactic medium \citep{hrl}, we consider two possible cases,
where either molecular cooling is effective or is not (atomic cooling
sets the threshold in the latter case). In either case, once the
minimum $V_c$ has been determined, the minimum halo mass in which gas
can cool efficiently is then \beq M_{\rm min}(z) = 9.4 \times 10^7
M_{\odot} \left( \frac{V_c}{16.5} \right)^3 \left( \frac{1+z}{10}
\right)^{-\frac{3} {2}} \left( \frac{\Omega_m h^2}{0.14} \right)^{-
\frac{1} {2}}\ .\label{eq:Mmin}\eeq

In addition, reionization heats the IGM, increasing pressure gradients
dramatically and preventing gas from accumulating in small halos
\citep{r86,e92,tw96,qke96, whk97,ns97,ki00, d04}. Specifically, we assume 
that after reionization gas continues to accrete only into halos above
a characteristic infall mass. For a small density fluctuation relative
to the cosmic mean density, gravity overcomes pressure in a region
above the Jeans mass \beq M_{\rm J} = 2.6\times 10^9 \left({\Omega_m
h^2\over 0.14}\right)^{- \frac{1} {2}} \left[\gamma\, \frac{0.61}{\mu}
\frac{10}{1+z} \frac{T}{10^4 K} \right]^{\frac{3} {2}} M_\odot\ , \eeq 
where $T$ is the temperature of gas at the cosmic density, $\gamma$
(which lies after reionization in the range 1--1.5) is its adiabatic
index, and $\mu$ is the mean molecular weight in units of the proton
mass. However, a halo in the process of formation is significantly
denser than the cosmic mean, so that even halos well below the Jeans
mass can accrete some gas, especially at high redshift. Based on the
above-mentioned numerical simulations, we adopt a minimum infall mass
of $M_{\rm J}/4$, which corresponds to a circular velocity of $\sim
35$ km/s. Note that it would be incorrect to use the filtering mass
\citep{gh98} as the cutoff mass, since the filtering mass describes the
cumulative time-averaged effect of pressure suppression while our
results depend on the instantaneous rate of gas infall into galactic
halos.

The Jeans mass depends on the thermal evolution of the IGM
\citep{hg97,Theuns}. In order to calculate it we assume that reionized 
gas is initially heated to $20,000$ K and subsequently evolves
according to \beq n \frac{d}{dt} \left( \frac{3}{2} k_B T \right) =
k_B T \frac{dn}{dt} + \Gamma - \Lambda \ , \label{eq:thermal} \eeq
where the number density of particles $n$ changes due to adiabatic
expansion, and the right-hand side includes the volume cooling rate
$\Lambda$ and heating rate $\Gamma$. The cooling rate includes the
contributions of single and dielectronic recombinations, collisional
excitations and ionizations, free-free emission, and Compton
scattering off the CMB. Assuming a highly-ionized plasma in ionization
equilibrium, the heating rate for each atomic species equals its
recombination rate times the average energy injected per
photoionization. We assume an ionizing spectrum of stars with
metallicity equal to $1/20$ of the solar value, distributed according
to the locally-measured IMF of \citet{scalo}. This implies that helium
is once-ionized simultaneously with hydrogen reionization, and that
later photoheating of recombining atoms results in an average energy
injection of 3.8 eV per recombination to H~I and 5.6 eV per
recombination to He~I [where we used \citet{Leith99}]. \citet{hg97}
showed that after reionization, the gas in hydrodynamic simulations
approximately follows a power-law temperature-density relation $T
\propto \rho^{\gamma -1}$ at densities up to $\sim 10$ times the mean
density. We estimate $\gamma$ by calculating the thermal evolution of
gas at the (evolving) cosmic mean density, and the temperature of gas
held at ten times the mean density.  

In general, star formation can be triggered when fresh gas is accreted
from the IGM (including during mergers) or from merger-induced
gravitational effects on gas that had previously accreted into
galaxies. In this paper we assume that the SFR in galaxies is
dominated by the fresh-gas mode, and comment below on the possibility
of a significant mode of star formation in previous-accreted gas. We
calculate the average accretion history of halos using the extended
Press-Schechter model. Given that a halo of mass $M_{\rm h}$ formed at
redshift $z_{\rm h}$, the fraction of its mass that had already
accreted into progenitor halos above $M_{\rm min}(z')$ at a given $z'
> z_{\rm h}$ is
\citep{bond91,lc93}
\beqa  & & F_{\rm acc}(M_{\rm h}, z_{\rm h}, M_{\rm min}(z'), z') =
\nonumber \\ & & \ \ \ \ \ \  {\rm 
erfc}\left[ \frac{\delta_c(z')- \delta_c(z_{\rm h})} {\sqrt{2
\left[S(M_{\rm min}(z')) - S(M_{\rm h}) \right]}} \right]\ ,
\label{eq:Facc} \eeqa 
where in general, $S(M)$ is the variance of
fluctuations in spheres of comoving radius containing mass $M$, and
$\delta_c(z)$ is the critical density threshold required for a region
to collapse by redshift $z$. We use the convention where the variance
is linearly-extrapolated to the present, while the critical threshold
varies with redshift. If we write $\delta_c(z) = \delta_0 (1+z)$, then
$\delta_0$ is approximately constant at high redshifts, and equals
1.28 for our assumed cosmological parameters.

The SFH of a halo depends on the assumed reionization redshift $z_{\rm
rei}$ of the surrounding region. For a halo containing a galaxy with
stellar mass $M_*$ at $z_{\rm h}$, the mean expected SFR at times
earlier than reionization is simply equal to \beq SFR(z') = M_*
\frac{d}{dt'} F_{\rm acc}(M_{\rm h}, z_{\rm h}, M_{\rm min}(z'),z')\ ,
\eeq where $t'$ is the age of the universe at redshift $z'$, and we
assumed a star formation efficiency that does not change with time. We
use this expression even though $M_{\rm min}$ changes with time, since
it only increases slowly with time and so we assume that gas that
accretes into a halo near $M_{\rm min}(z')$ at some $z'$ is likely to
continually merge into larger halos and thus be in a halo above
$M_{\rm min}(z)$ at all $z < z'$. At reionization, however, the
minimum mass for accreting new gas jumps suddenly to $\frac{1}{4}
M_{\rm J}(z_{\rm rei})$, and a large fraction of the gas that had
previously accreted remains for some time in halos less massive than
this minimum infall mass. At a given redshift $z'$ after reionization,
the fraction of the halo gas that has accumulated since reionization
equals the fraction that is in halos above $\frac{1}{4}M_{\rm J}(z')$
and that had not yet accreted into galaxies by $z_{\rm rei}$. For each
possible progenitor mass at $z'$, we apply equation~(\ref{eq:Facc}) to
find the fraction of the progenitor's mass that still lay outside
galactic halos at reionization. Note that this fraction is independent
of the final halo mass $M_{\rm h}$ due to the random walk statistics
of the extended Press-Schechter model \citep{bond91}. For progenitors
of a given mass $M$ at redshift $z'$, infall of new gas is due to both
the increasing number of progenitors with time and the increase with
time of the fraction of gas in the progenitors that did not already
form stars before $z_{\rm rei}$. Thus, in this case \beqa SFR(z') & =
& M_* \frac{d}{dt'} \int_{\frac{1}{4} M_{\rm J}(z')}^{M_{\rm h}} dM\,
\left[ \frac{d} {dM} F_{\rm acc} (M_{\rm h}, z_{\rm h}, M, z') \right]
\nonumber \\ & & \times \left\{1- F_{\rm acc}(M, z', M_{\rm
min}(z_{\rm rei}),z_{\rm rei}) \right\}\ , \eeqa where $M_{\rm
min}(z_{\rm rei})$ is calculated from eq.~(\ref{eq:Mmin}) assuming
either atomic or molecular cooling sets the minimum mass.

Our approach in this paper to reionization is not to try to
theoretically predict the absolute redshift at which it ends, since
this depends strongly on the average escape fraction of photons from
galaxies, a number that is highly uncertain. We instead focus on
observational constraints on the reionization redshift as well as
theoretical predictions of the spatial {\it fluctuations}\/ in the
reionization history. In particular, we define a redshift shift
$\Delta z$ which is the difference between the local redshift at which
a given region is reionized and the redshift of complete global
reionization. Regions in which galaxies form earlier than the cosmic
average have a positive $\Delta z$. The expected distribution of
$\Delta z$ values can be predicted for a given reionization redshift,
as long as the escape fraction from galaxies has roughly the same
overall average value in different regions. We focus on these redshift
fluctuations because they may be directly observable, as we argue
below.

Massive halos are rather rare at high redshift, but current telescopes
can only detect the very brightest galaxies from those early times. At
high redshift there are large fluctuations in the number density of
galaxies in various regions \citep{BLflucts}. The extended
Press-Schechter model predicts that galactic halos will form earlier
in regions that are overdense on large scales \citep{k84, b86, ck89,
mw96}, since these regions already start out from an enhanced level of
density, and small-scale modes need only supply the remaining
perturbation necessary to reach the critical density for
collapse. Thus, an observed bright galaxy is likely to naturally flag
a region with a large-scale overdensity. In such a region, galaxy
formation and thus reionization will proceed faster than in the
universe as a whole. In particular, within the extended
Press-Schechter model \citep{bond91}, if the cosmic fraction of gas
accumulated in galaxies reaches a certain value at redshift $z$, then
it will reach the same value at a different redshift $z+\Delta z$ in a
box containing mass $M_{\rm box}$ with a mean overdensity
$\bar{\delta}_{\rm box}$. At high redshifts ($z > 3$), this shift in
redshift is \citep{BLflucts}
\begin{equation} \Delta z = \frac{\bar{\delta}_{\rm box}}
{\delta_0} - (1+z) \times \left[ 1 - \sqrt{1 - \frac{S(M_{\rm box})}
{S(M_{\rm min})}}\ \right]\ . \label{eq:Deltaz}
\end{equation}

To calculate the number density of halos above some mass in a given
region, we begin with the cosmic mean halo mass function of
\citet{shetht99} which fits numerical simulations more accurately
\citep{jenkins} than the original model of \citet{ps74}. We
then adjust the halo distribution as a function of density in a region
based on the prescription of \citet{BLflucts}, in which the extended
Press-Schechter model is used for the relative adjustment factor; we
showed there that this hybrid model fits a wide range of
simulations. If the expected number of halos above some mass in a
given region is $\bar{n}$, then the probability of finding at least
one such halo in such a region is given by Poisson statistics: $P(\ge
1)=1-{\rm exp}[-\bar{n}]$. Furthermore, since $\bar{n}$ is a function
of the overdensity $\bar{\delta}_{\rm box}$ of the region, the
probability of the region having a given $\bar{\delta}_{\rm box}$ can
be calculated as the probability of having an overdensity
$\bar{\delta}_{\rm box}$ [which is given by a Gaussian with variance
$S(M_{\rm box})$] times $P(\ge 1)$ for the given $\bar{\delta}_{\rm
box}$, divided by $P(\ge 1)$ for the cosmic mean halo mass function.

These probability distributions of $\Delta z$ also imply a large
scatter in the reionization redshift in various regions. Once ionizing
photons escape from the immediate vicinity of the sources, they
encounter a mostly low-density IGM, where the dominant factor limiting
reionization is the need to produce one ionizing photon per hydrogen
atom \citep{mhr00,g00}. If we assume a fixed efficiency for the
production of ionizing photons in galactic halos, then the reionized
fraction in any region is thus proportional to the total gas fraction
in galaxies within that region. This assumption is in fact fairly
accurate even with recombinations included, as long as the number of
times each hydrogen atom must be reionized is roughly the same in
different regions. Indeed, the recombination rate is not expected to
vary greatly in different regions, since the fluctuation in the mean
density of such regions is fairly small. For example, a 10 comoving
Mpc box at redshift 7 has density fluctuations of $\sigma \sim
0.2$. Recombinations can be included in a future refinement of our
model.

Numerical simulations in small boxes \citep{g00} suggest that
reionization occurs locally as a fast transition, as we have assumed
for the progenitors of each galaxy, while the large-scale
galaxy-number fluctuations mentioned above imply that globally
reionization occurs at different times in different
regions\citep{BLflucts}. Rare over-dense regions can be assumed to
reionize themselves as though they were isolated from the rest of the
universe, since they are most likely to fully reionize before their
lower-density surroundings. Thus, the $\Delta z$ calculated above
yields an accurate estimate for the earlier reionization redshift of
such regions, as long as $\Delta z$ is positive and significantly
large. We assume this is the case for regions that are at least a
$+1-\sigma$ fluctuation, which means that they form more galaxies than
$84\%$ of other such regions in the universe. For voids, on the other
hand, while $\Delta z$ of reionization can come out negative in our
model, the real $\Delta z$ in the universe cannot be negative for any
region; as soon as the universe contains enough ionizing photons to
fully reionize, all regions are indeed reionized, including the
underdense regions which are reionized by excess photons from the
denser surrounding regions. Thus, we adjust the probability
distribution of $\Delta z$ for reionization, for all $\Delta z$
smaller than that corresponding to a $+1-\sigma$ fluctuation, by a
constant factor so that the total probability of having $\Delta z > 0$
is unity.

These arguments suggest that the SFHs of high-redshift galaxies can be
used to study the spatial correlation function of reionization. In a
sufficiently small region, all galaxies will show in their SFH
evidence for approximately the same reionization redshift which
characterizes that particular region. Galaxies separated by a large
distance will in general show a significant variation in reionization
redshifts, where this variation will be strongly correlated with the
different environments of the galaxies. In the quantitative results
below, we adopt a distance of 10 comoving Mpc as the fiducial size of
a region with a uniform reionization history, based on analytical
models of the distribution of bubble sizes during reionization
\citep{FurHII1,WL04}.

\section{Results}

\begin{figure}
\includegraphics[width=84mm]{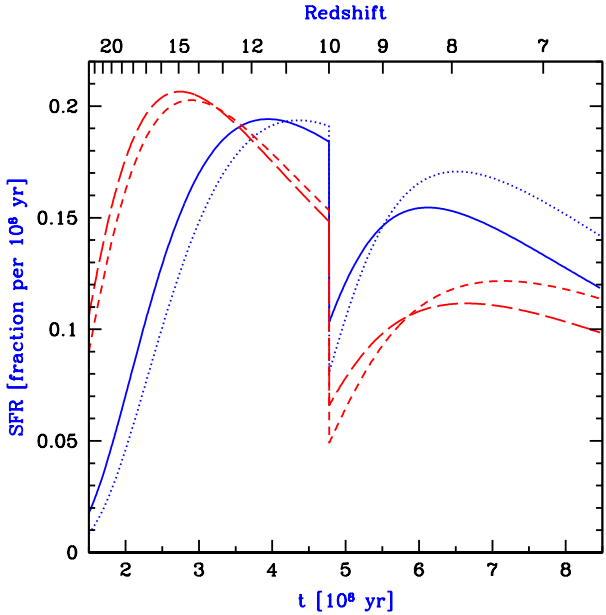}
\caption{SFHs of redshift 6.5 galaxies vs.\ cosmic age or redshift. 
We show the fractional SFR summed over all progenitors that form
stars. Assuming $z_{\rm rei}=10$, we consider atomic cooling for the
progenitors of $M_{\rm h}=10^{13} M_{\odot}$ halos (solid curve) or
$M_{\rm h}=10^{11} M_{\odot}$ halos (dotted curve), or molecular
cooling for the progenitors of $M_{\rm h}=10^{13} M_{\odot}$ halos
(long-dashed curve) or $M_{\rm h}=10^{11} M_{\odot}$ halos
(short-dashed curve).}
\label{fig:SFHgen}
\end{figure}

\begin{figure}
\includegraphics[width=84mm]{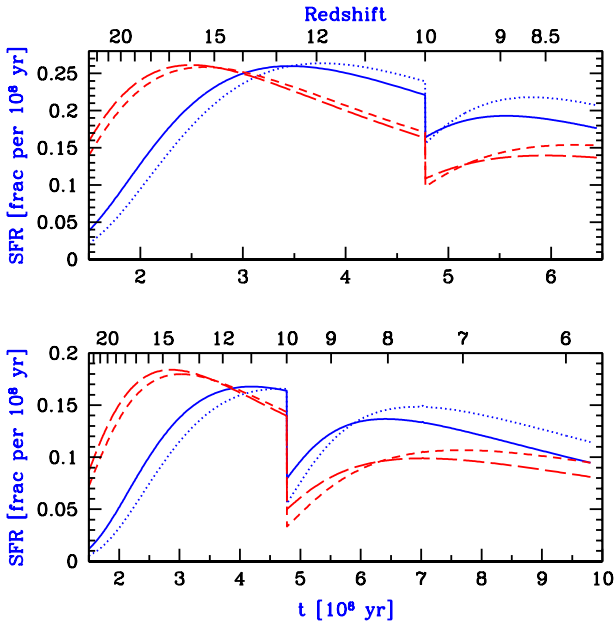}
\caption{Same as Figure~\ref{fig:SFHgen}, but for redshift 8 galaxies 
(top panel) and for redshift 5.8 galaxies (bottom panel). Note that in
each panel the $x$-axis shows cosmic age (bottom) and redshift (top).}
\label{fig:SFHgen2}
\end{figure}

\begin{figure}
\includegraphics[width=84mm]{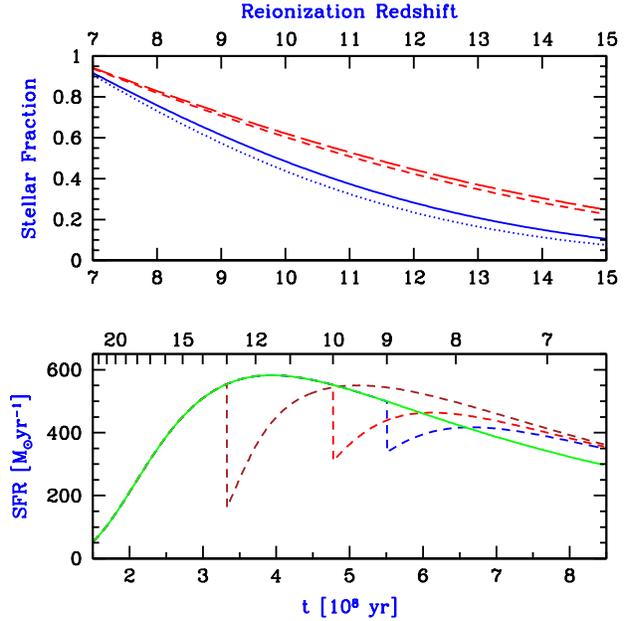}
\caption{{\it Bottom panel}: Same as Figures~\ref{fig:SFHgen} and 
~\ref{fig:SFHgen2} (with the $x$-axis shown as both time and
redshift), but for the case of a $10^{13} M_{\odot}$ halo and atomic
cooling, assuming no reionization (solid curve) or $z_{\rm rei}=9$,
10, or 13 (dashed curves). We show the mean expected SFR in absolute
units, assuming a $20\%$ star formation efficiency. {\it Top panel}:
Fraction of stars that formed prior to reionization vs.\ $z_{\rm
rei}$, for a halo observed at $z=6.5$. We consider atomic cooling for
halos $M_{\rm h}=10^{13} M_{\odot}$ (solid curve) or $M_{\rm
h}=10^{11} M_{\odot}$ (dotted curve), and molecular cooling also for
$M_{\rm h}=10^{13} M_{\odot}$ (long-dashed curve) or $M_{\rm
h}=10^{11} M_{\odot}$ (short-dashed curve).}
\label{fig:Fraction}
\end{figure}

The reionization signature we predict is a generic feature, and we
illustrate it in this section for a wide range of possible
galaxies. For example, Figure~\ref{fig:SFHgen} shows the mean expected
SFHs of redshift 6.5 galaxies. As detailed in the previous section, we
assume that stars form in halos above a minimum cooling mass before
reionization, and in halos above a minimum infall mass after
reionization. We consider a pre-reionization minimum mass set by
requiring atomic cooling or molecular cooling. We find that a massive
$z=6.5$ galaxy should generically show a double-peaked SFH, with one
early peak due to star formation in low-mass progenitors before
reionization, and a much later peak due to gas infall into more
massive progenitors. The same qualitative features are expected even
when we vary the observed redshift (see Figure~\ref{fig:SFHgen2}) or
the reionization redshift (see the bottom panel of
Figure~\ref{fig:Fraction}). For a given observed galaxy, if
reionization occurred earlier, then its signature on the SFH is more
difficult to observe, for two reasons. First, a smaller fraction of
the observed stars is expected to have formed prior to reionization
(see the top panel of Figure~\ref{fig:Fraction}); and second, the
early peak in this case lies more than a few hundred million years in
the past, allowing only low-mass stars to survive and making it
difficult to pin down the age of this episode of star formation.

\begin{figure}
\includegraphics[width=84mm]{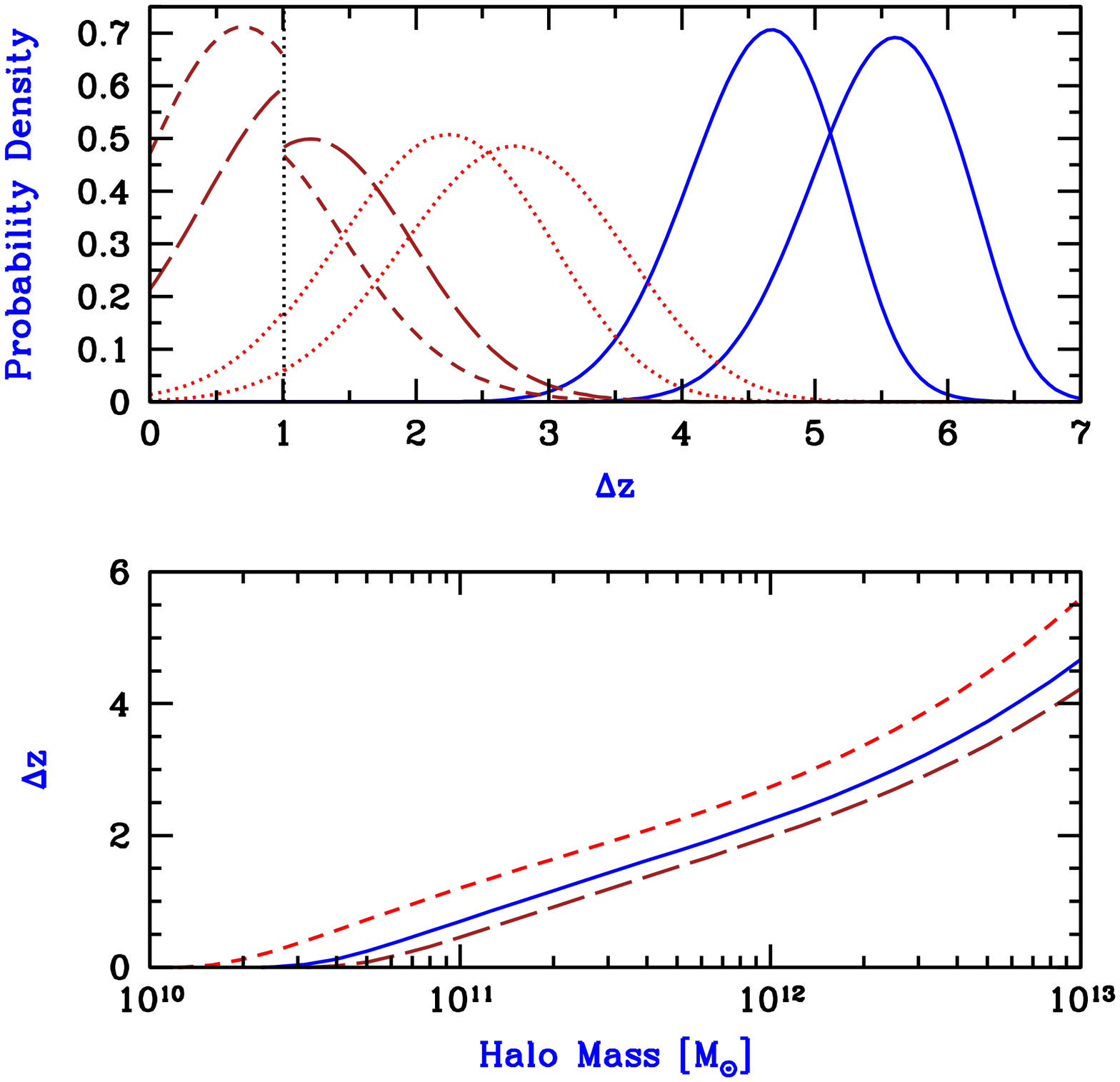}
\caption{{\it Bottom panel}: Most probable redshift shift vs.\ halo mass, 
for a halo observed at $z=5.8$ (long-dashed curves), 6.5 (solid
curves), or 8 (short-dashed curves). In each case, we show $\Delta z$
for reionization of a 10 Mpc box containing the halo, relative to
complete reionization of the universe. {\it Top panel}: Normalized
probability distribution of $\Delta z$. We consider halos $M_{\rm
h}=10^{13} M_{\odot}$ (solid curves) or $M_{\rm h}=10^{12} M_{\odot}$
(dotted curves), considering (from left to right) a halo observed at
$z=6.5$ or at $z=8$. For the $M_{\rm h}=10^{11} M_{\odot}$ case, we
also consider $z=6.5$ (short-dashed) or $z=8$ (long-dashed). Note that
to the left of the dotted vertical line (which indicates a $+1-\sigma$
fluctuation), the total probability in each case is fixed but its
precise distribution is uncertain (see \S~2).}
\label{fig:Deltaz}
\end{figure}

\begin{figure}
\includegraphics[width=84mm]{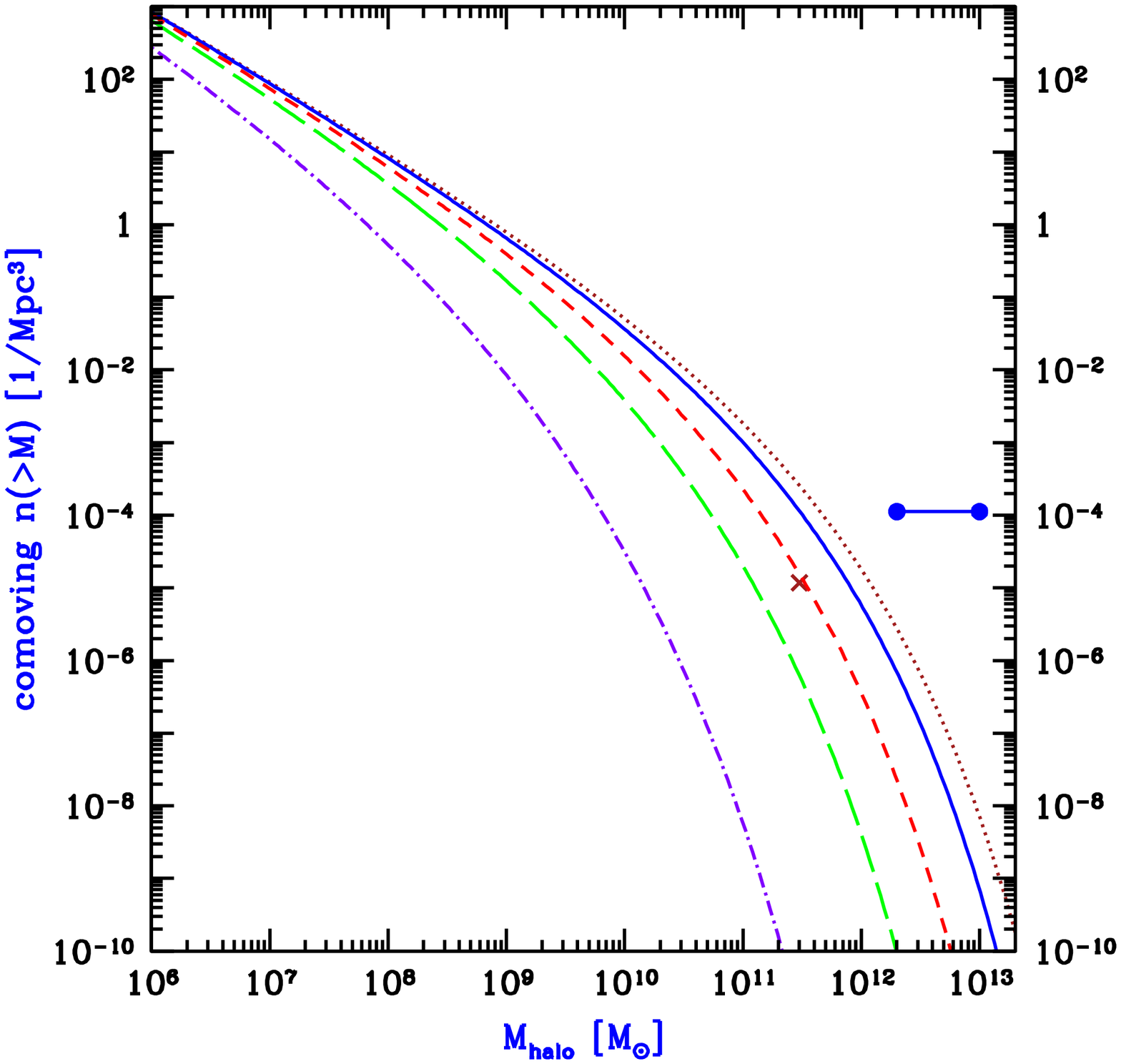}
\caption{Cosmic mean number density of halos versus halo mass. We
show the comoving number density at $z=5.8$ (dotted curve), 6.5 (solid
curve), 8 (short-dashed curve), 10 (long-dashed curve), and 15
(dot-dashed curve). Also shown are the observed abundances of massive
$z=5.8$ galaxies using \citet{eyles05} and assuming a $20\%$ star
formation efficiency ($\times$), and the range (two connected circles)
corresponding to an efficiency of $20\%$--100$\%$ for the massive
$z=6.5$ galaxy from
\citet{z6p5a}.}
\label{fig:number}
\end{figure}

Current instruments are limited to detecting the brightest
high-redshift galaxies, which should typically reside in the most
massive halos. Such halos are far more likely to form in overdense
regions, where galaxy formation generally, and reionization in
particular, should occur earlier than in the universe as a
whole. Figure~\ref{fig:Deltaz} shows that substantial values of
$\Delta z$ up to $\sim 7$ are possible, depending on the halo mass and
the redshift at which it is observed. Note that the value of $\Delta
z$ is essentially independent of the reionization redshift and of the
precise value of the minimum mass, since for both atomic and molecular
cooling, $S(M_{\rm min}) \gg S(M_{\rm box})$ in eq.~(\ref{eq:Deltaz}).
Increasingly rare halos should show larger values of $\Delta z$, a
trend that should be seen observationally once large-area surveys
determine the abundance of rare, massive
halos. Figure~\ref{fig:number} shows the mean abundance of halos at
several redshifts \citep{shetht99}.

\section{Application to Particular Galaxies and Discussion}

\citet{z6p5a} have recently used deep multi-waveband photometry to 
place the galaxy HUDF-JD2 potentially at a redshift $z \sim 6.5$. They
estimated a stellar mass of $6\times 10^{11}M_\odot$, for a Salpeter
stellar IMF which places most of the mass in stars below a solar
mass. Fixing the contribution of the 1--$3M_\odot$ stars which
dominate the observed infrared emission, we convert to the
locally-measured IMF of \citet{scalo} and find a total stellar mass of
$3\times 10^{11}M_\odot$. This implies a minimum host halo of total
mass $2\times 10^{12}M_\odot$, for a maximum possible ($100\%$)
efficiency of formation of stars from baryons. Efficiencies near $\sim
10\%$ are more typically assumed, so we illustrate results for an
intermediate value of $20\%$ which yields a $10^{13}M_\odot$
halo. \citet{eyles05} have also analyzed two high-redshift galaxies
with spectroscopic redshifts $z=5.78$ and $z=5.83$, with stellar
masses $\ga 10^{10} M_{\odot}$ [for the \citet{scalo} IMF], which for
a $20\%$ efficiency implies host halos of mass $3\times
10^{11}M_\odot$ for each galaxy.

Galaxies similar to those observed should generally show a clear
double-peaked SFH. For a $10^{13} M_\odot$ galactic halo at $z=6.5$, a
majority of the stars are expected to have formed prior to
reionization if reionization occurred at $z_{\rm rei} < 9.9$ or
$z_{\rm rei} < 11.3$, assuming atomic or molecular cooling,
respectively (see Figure~\ref{fig:Fraction}). These redshifts
correspond to look-back times of 363 and 443 million years,
respectively, with respect to the cosmic age of 847 million years at
$z=6.5$. A $3\times 10^{11}M_\odot$ halo observed at $z=5.8$ should
have its SFH dominated by pre-reionization stars if $z_{\rm rei} <
8.9$ (atomic cooling) or $z_{\rm rei} < 10.4$ (molecular
cooling). Indeed, signatures of early star formation were found in the
galaxies cited above. \citet{eyles05} found that most of the stars in
the two $z=5.8$ galaxies had formed at $z \ga 7.5$. Similarly,
\citet{z6p5a} found that the $z=6.5$ galaxy had formed most of its
stars at $z \ga 9$, and could have reionized its surrounding region
even earlier, at $z \ga 10$ \citep{z6p5b}.

If these signatures do indeed correspond to local reionization of the
regions surrounding these galaxies, then they could have important
implications for cosmic reionization. However, in analyzing these
implications we must account for the fact that the galaxies were most
likely found in overdense regions that reionized earlier than did the
rest of the universe. Indeed, the 10 comoving Mpc region surrounding
the massive $z=6.5$ galaxy most likely reionized at a $\Delta z = 4.7$
earlier than the rest of the universe, i.e., at $z \ga 11$. The
observed $z=5.8$ galaxies correspond to halos that are not nearly as
rare, but still are most likely in regions that reionized early by a
$\Delta z = 1.2$, i.e., at $z \ga 7.5$. Thus, the signatures of
reionization observed in these galaxies only imply an early
reionization of the surrounding regions, but do not push cosmic
reionization earlier than the already established $z \ga 6.5$ (see
\S~1).

If these galaxies indeed lie in atypical, overdense regions, then
consistency implies that such massive galaxies must be extremely
rare. The two $z=5.8$ galaxies were detected in an effective volume of
$\sim 1.7 \times 10^5$ comoving Mpc$^3$ \citep{sbm03}, while the
$z=6.5$ galaxy was found in the Hubble Ultra Deep Field, which
(adopting an effective redshift range of 6.2--6.8 for the search)
corresponds to $\sim 9 \times 10^3$ comoving Mpc$^3$. Although the
number of galaxies is small and large fluctuations are expected, we
can estimate the resulting abundances of these galaxies (see
Figure~\ref{fig:number}). The $z=5.8$ galaxies have a lower abundance
than expected, by a factor of $\sim 20$. However, the observed number
of two galaxies represents a lower limit, since there are other
galaxies in the field with undetermined redshifts or stellar
masses. Furthermore, if the star formation efficiency in these
galaxies is only $\sim 5\%$ rather than $20\%$, then their abundance
is similar to that expected. However, the $z=6.5$ galaxy is extremely
massive and was found in a very small field. For the $20\%$ efficiency
that we have adopted, the a posteriori probability of finding this
galaxy is $\sim 6 \times 10^{-6}$, making it rather unlikely that the
claimed high redshift is accurate. If this galaxy represents the
result of an unusual star formation episode which was characterized by
a near-perfect efficiency, then finding such a $2 \times 10^{12}
M_\odot$ halo represents a $\sim 10^{-3}$ likelihood. Further
observations of larger fields are required to determine whether this
detection may have been a stroke of luck. Note, however, that even a
small change in the halo mass would change the detection probability
significantly, because of the exponential cutoff of the halo mass
function. Thus, the probability is also sensitive to any uncertainties
in the estimated stellar mass.

We caution that the SFHs we predict in this paper may not be entirely
accurate. First, even in a small region from which a single observed
galaxy arises, the suppression of star formation builds up somewhat
gradually as the progenitors reionize their surroundings. In addition,
if significant star formation is triggered during mergers in
previously-accreted gas then this will somewhat weaken the signature
of reionization, since infall of fresh gas is immediately suppressed
by the increased pressure while mergers continue among progenitors
that had accreted gas prior to reionization. We expect, however, that
infall will be dominant at high redshift, since around reionization
most of the gas of an eventual large galaxy lies in low-mass
progenitor halos (typically with mass of order $10^8 M_{\odot}$) in
which even a few supernovae can easily disperse much of the gas. Thus,
although more detailed models of feedback should be explored, we
expect the initial burst of star formation in fresh infalling gas to
be the dominant mode of star formation at high redshift. It is also
important to note that we have calculated in this paper mean expected
SFHs, while there should be significant variation in the merger
histories of individual galaxies (e.g., see the top panel of
Fig. \ref{fig:Deltaz}). Detailed models and numerical simulations can
be used to more accurately quantify our predictions. However,
regardless of the fine details, we emphasize that the signature we
predict is a generic, qualitative feature that should be clearly
observable in a majority of high-redshift galaxies.

In conclusion, observations of high-redshift galaxies should always be
fitted with SFHs consisting of at least two separate episodes of star
formation. If the signature of reionization is seen in the SFHs of
many individual galaxies, then the variation of the reionization
redshift in different regions and its spatial correlation function can
be studied quantitatively. In particular, a consistent picture should
emerge where overdense regions both reionize earlier and contain a
higher abundance of massive galaxies than the rest of the universe.

\section*{Acknowledgments}
R.B. is grateful for the kind hospitality of the {\it Institute for
  Theory \& Computation (ITC)} at the Harvard-Smithsonian CfA, and
acknowledges support by Harvard university and the Israel Science
Foundation grant 28/02/01. This work was supported in part by NASA
grants NAG 5-1329 and NNG05GH54G (for A.L.).

\bsp

\label{lastpage}

\end{document}